\documentclass[preprint]{jfm} %% "referee" for double spaced
\usepackage{graphicx}
\usepackage{epsfig}
\usepackage{bm}
\usepackage{amsmath}
\usepackage{amssymb}
\usepackage{wasysym}
\usepackage{color}
\usepackage{xcolor}
\usepackage{epstopdf}
\usepackage{upgreek}
\usepackage{subfloat}
\usepackage{graphicx}% Include figure files
\usepackage{dcolumn}% Align table columns on decimal point
\usepackage{bm}% bold math
\usepackage{latexsym}
\usepackage{subfloat}
\usepackage[amssymb]{SIunits}
\usepackage{subfig}
\usepackage{psfrag}
\usepackage{xcolor}
\usepackage{natbib}
\usepackage{amsmath}
\usepackage{comment}
\usepackage[normalem]{ulem}
\graphicspath{{Figures//}}
\usepackage[colorlinks=false, bookmarks=true]{hyperref}

\title[Energy spectra in turbulent bubbly flows]{Energy spectra in turbulent bubbly flows} 
\author[Prakash, V. N. et al.]
{Vivek  N. Prakash$^{1,2}$,
J. Mart\'inez Mercado$^{1}$, %\break
Leen van Wijngaarden$^{1}$,
E. Mancilla$^{1,3}$, 
Y.  Tagawa$^{1,4}$,
Detlef Lohse$^{1,5}$,  and Chao Sun$^{1,6}$\ns
\thanks{Corresponding author. E-mail: c.sun@utwente.nl} 
}

\affiliation{$^1$Physics of Fluids Group, Faculty of Science and Technology, J.M. Burgers Center for Fluid Dynamics, University of Twente, P.O. Box 217, 7500 AE  Enschede, The Netherlands\\[\affilskip]
$^2$Department of Bioengineering, Stanford University, Stanford, California 94305, USA\\[\affilskip]
$^3$Instituto de Investigaciones en Materiales, Universidad Nacional Aut\'onoma de M\'exico, M\'exico Distrito Federal 04510, Mexico\\
$^4$Department of Mechanical Systems Engineering, Tokyo University of Agriculture and Technology, 1848588, Koganei-city, Tokyo, Japan
\\
$^5$Max Planck Institute for Dynamics and Self-Organization, D-37077 G$\ddot{o}$ttingen, Germany\\
$^6$Center for Combustion Energy, and Department of Thermal Engineering, Tsinghua University, Beijing 100084, China \\
[\affilskip]}
\date{\today}

\begin{document}
\maketitle

\begin{abstract} 
We conduct experiments in a turbulent bubbly flow to study the nature of the transition between the classical $-$5/3  energy spectrum scaling for a single-phase turbulent flow and the $-$3 scaling for a swarm of bubbles rising in a quiescent liquid and of bubble-dominated turbulence. The bubblance parameter \cite[]{lance,judith}, which measures the ratio of the bubble-induced kinetic energy to the kinetic energy induced by the turbulent liquid fluctuations before bubble injection, is often used to characterise the bubbly flow. We vary the bubblance parameter from $b = \infty$ (pseudo-turbulence) to $b = 0$ (single-phase flow) over 2-3 orders of magnitude ($0.01 - 5$) to study its effect on the turbulent energy spectrum and liquid velocity fluctuations. 
The probability density functions (PDFs) of the liquid velocity fluctuations show deviations from the Gaussian profile for $b > 0$, i.e. when bubbles are present in the system. The PDFs are asymmetric with higher probability in the positive tails. 
The energy spectra are found to follow the $-$3 scaling at length scales smaller than the size of the bubbles for bubbly flows. This $-$3 spectrum scaling holds not only in the well-established case of pseudo-turbulence, but surprisingly in all cases where bubbles are present in the system ($b > 0$). Therefore, it is a generic feature of turbulent bubbly flows, and the bubblance parameter is probably not a suitable parameter to characterise the energy spectrum in bubbly turbulent flows. The physical reason is that the energy input by the bubbles passes over only to higher wave numbers, and the energy production due to the bubbles can be directly balanced by the viscous dissipation in the bubble wakes as suggested by \cite{lance}. In addition, we provide an alternative explanation by balancing the energy production of the bubbles with viscous dissipation in the Fourier space. 

\end{abstract}

%%%----------------------------------------INTRODUCTION ------------------------------------------------%%%
\section{Introduction}
\label{sec:Introduction}
Turbulent bubbly flow has important industrial applications such as in chemical industries and steel plants~\cite[]{DeckwerBook}. A fundamental understanding of the influence of bubbles on turbulence
is crucial for better designs and optimal utilization of resources~\cite[]{magnaudet2000,ern2012}. 
In this work we investigate the  effect of bubbles on properties such as velocity fluctuations and energy spectra of turbulent flows.  In turbulent bubbly flows the energy input in the liquid  comes from both the bubbles and from external forcing. Depending on this source of energy input we can distinguish different regimes. Various parameters  have been used to characterize these regimes. One of these is the ``bubblance" b  \cite[]{lance,judith} defined as,
\begin{equation}
b=\frac{1}{2} \frac{\alpha U^2_r}{u^{\prime2}_0}, \label{eq:bubblance}
\end{equation}
 where, $\alpha$ is the bubble concentration (void fraction), $U_r$ is the  bubble rise velocity in still water, and $u^{\prime}_0$ is the typical turbulent liquid fluctuation in the absence of bubbles. 

We see that the case  b=0  represents  single-phase  turbulence  and b $= \infty$ the situation where the fluctuating velocities are purely caused by  bubbles.  The latter regime is often called pseudo-turbulence, which has been extensively investigated experimentally.
A few early studies~\cite[e.g.][]{cui,mudde} reported a Kolmogorov type of spectral density behavior, a dependence of the wave number k as $k^{-5/3}$. However, recent work \cite[]{julian2,mendez2013,riboux2013} clearly established that the spectral density in pseudo-turbulence behaves as $k^{-3}$. In fact, the {\it{-3}} spectrum scaling is found to be robust even if the bubble size is changed, or if higher viscosity liquids are used instead of water. The recent study by \cite{mendez2013} suggests that the specific details of the hydrodynamic interactions among bubbles do not influence the way in which the pseudo-turbulent fluctuations are produced. The current understanding is that  the bubble-induced turbulence mainly results from the bubble wakes. The importance of the bubble wakes on the {\it{-3}} spectrum scaling has also been established using numerical simulations by comparing the spectrum scalings between point-like bubble simulations~\cite[]{mazzitelli} and fully-resolved simulations of freely rising deformable bubbles~\cite[]{roghair}, with the former one giving {\it{-5/3}} due to the absence of wakes, and the latter fully-resolved simulations giving {\it{-3}} as the spectral scaling exponent.

Previous work has mainly been concerned with the extreme values of the $b$ parameter, i.e. either pseudo-turbulence ($b = \infty$) or single-phase turbulence (b = 0). Our focus in this paper is to study what happens \textit{in between these extremes} $b = \infty$ and $b = 0$ as the energy spectrum scaling and the liquid velocity fluctuation statistics are not well known for large ranges of intermediate $b$. 
In this paper, we thus want to systematically analyse the flow as a function of the $b$ parameter between the cases of single-phase turbulence ($b = 0$), turbulence with some bubbles ($0<b<5$). The $b$ parameter is varied over 2-3 orders of magnitude, namely from $0.01$ to $5$, and the pseudo turbulent case $b = \infty$ is also included. 
Pioneering experiments  in  turbulent bubbly flow were made by~\cite{lance} using hot-wire and Laser Doppler Anemometry (LDA). Their experiments did not cover a broad range of b. They found that with increasing $\alpha$ the Kolmogorov $k^{-5/3}$ spectral density is progressively substituted by a $k^{-8/3}$ behavior. \cite{lance} 
attributed this to the wakes of the bubbles. Here eddies are produced which are so rapidly dissipated that they do not take part in the energy cascade. Omitting then the transport term in the energy balance in k space, they concluded to a $k^{-3}$ behavior on dimensional ground. Referring to the definition of b in eq.~\ref{eq:bubblance}, one would expect a different spectral scaling depending on the energy input. For b $\ll$
1, the turbulent fluctuations would be dominant and the spectrum exponent is consequently close to $-5/3$. When b$\gg$1, bubble-induced fluctuations would dominate and the exponent would be close to $-3$. 
In the present work the parameter b is varied  by adding external turbulence of varying intensity which is produced using an active grid.

In the next section, we describe the experimental setup, tools and methods used. This is followed by the results section where we describe our findings for the liquid velocity fluctuations and energy spectra. We provide an interpretation of our results in the discussion section, where we also summarise our work.

%%%----------------------------------------EXPERIMENTS ------------------------------------------------%%%
\section{Experiments}

\subsection{\label{sec:level2}Experimental setup}

\begin{figure}
\centering
\subfloat{\includegraphics[width=7cm]{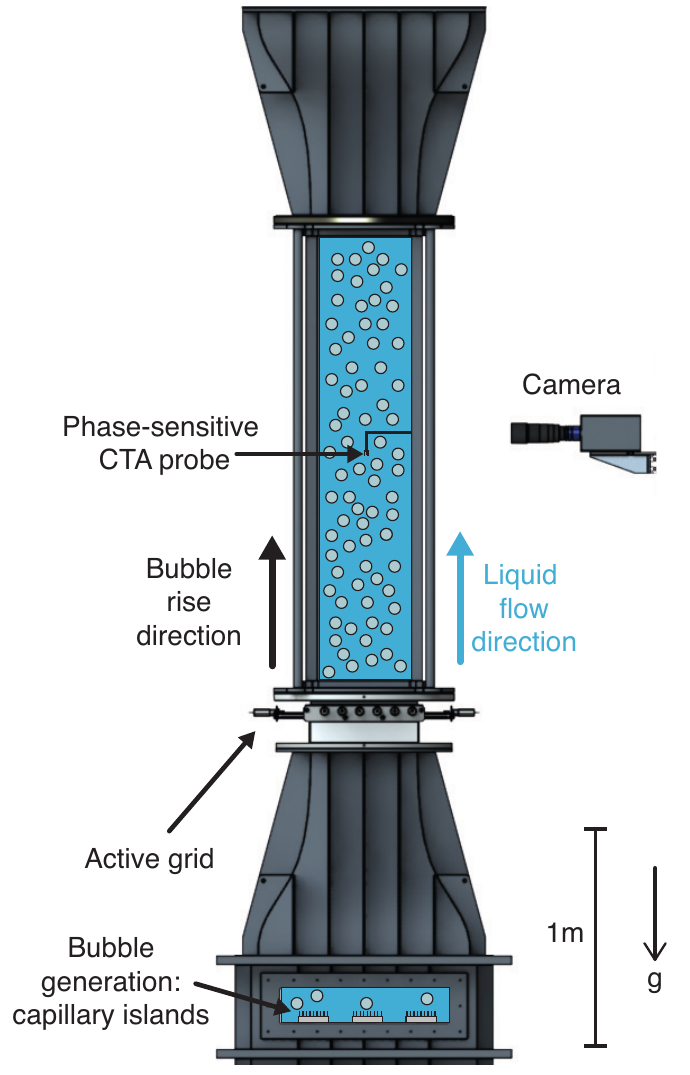}}
\caption{The Twente Water Tunnel (TWT) facility: A vertical multiphase water tunnel where homogeneous and isotropic turbulence is generated by an active-grid. Air is blown through capillary islands located below the measurement section to generate bubbles. The bubble rise direction and liquid flow are both in the upward direction, and the phase-sensitive CTA probe measures the liquid velocity fluctuations.}
\label{fig:expt_schem}
\end{figure}

The experiments are carried out in the Twente Water Tunnel (TWT) facility, which is an $8m$-high vertical water tunnel (see Figure~\ref{fig:expt_schem}). The measurement section of the TWT (dimensions: $2m \times 0.45m \times 0.45m$) is made of transparent glass to provide optical access for flow visualization and measurements. We place the phase-sensitive CTA  (hot-film) probe in the centre of this measurement section, more details on this technique are discussed in the next section. An active-grid is used to generate nearly homogeneous and isotropic turbulent flow in the liquid phase and it is placed below the test section~(\cite{Poorte2002,prakash1,prakash2}). Air bubbles are generated by blowing air through islands of capillary needles that are located below the measurement section. A U-tube setup mounted in the measurement section is used to measure the gas void fraction $\alpha$ (see~\cite{judith,julian2} for more details). The bubbles pass through the active-grid, rise through the measurement section and eventually escape through an open vent at the top of the TWT. The liquid mean flow is driven by a pump which recirculates the water throughout the TWT. The bubbles rise along with the upward mean flow in the measurement section; the system is thus a co-flowing turbulent upward bubbly flow.

In our experiments, we vary the bubble diameter by changing the inner diameter of the capillary needles in the bubble generating islands. Two different needles are used in our experiments: air bubbles of diameter $3$-$5$ mm are generated using capillary needles of inner diameter of 500 $\mu$m, and $2$-$4$ mm bubbles are produced using the needles with inner diameter of 120 $\mu$m. We classify our experiments into two sets based on the bubble diameter - experiments with bubbles of diameter $3$-$5$ mm belong to set 1, and experiments with bubbles of diameter $2$-$4$ mm are referred to as set 2 (see Table ~\ref{tab:kd}). The purpose of exploring different bubble diameters in sets 1 and 2 is to test for size effects and to achieve a wider range of $b$ values.

In the present experiments, the $b$ parameter is varied by changing: (i) the volume flow rate of air (i.e. equivalent to changing the gas void fraction $\alpha$) through the capillary islands, (ii) the magnitude of the mean flow speed of water in the upward direction (to effectively change the turbulence intensity $u^{\prime}_0$). In equation~\ref{eq:bubblance}, $U_r$ is the typical rise velocity of bubbles (in still water). We assume the value of the rise velocity of bubbles in still water, $U_r \approx 23$ cm/s, which is reasonable because it holds for the range of bubble diameters for both data sets (2-5 mm) considered in this study (see Fig 7.3 in~\cite{clift}).

\begin{table}
 \begin{center}
\def~{\hphantom{0}}

\begin{tabular}{cccc|cccc}
 \multicolumn{8}{c}{Set 1 (3-5 mm bubbles)~~~~~~~~Set 2 (2-4 mm bubbles)} \\
 \hline
      \bf{b}&\textbf{$\alpha$}&$u^{\prime}_0$&$U_{l}$&\bf{b}&$\alpha$&$u^{\prime}_0$&$U_{l}$\\ 
       ~&$\%$&cm s$^{-1}$&cm s$^{-1}$~&~&$\%$&cm s$^{-1}$&cm s$^{-1}$\\ 
       \hline
       $\infty$  & 2 & 0 & 0 & $\infty$  & 2  & 0 & 0 \\
       $\infty$  & 1.17 & 0& 0 & $\infty$  & 1.5 & 0 & 0 \\
       4.13   & 1 & 0.8 & 10 & $\infty$  & 1  & 0 & 0 \\
       2.06  & 0.5 & 0.8 & 10 & $\infty$  & 0.8 & 0 & 0 \\
       1.03   & 1 & 1.6 & 20 & $\infty$  & 0.5 & 0 & 0 \\
       0.78   & 0.75 & 1.6 & 20 & 4.13 & 1 & 0.8 & 10 \\ 
       0.52   & 0.5 & 1.6 & 20 & 2.06  & 0.5 & 0.8  & 10 \\
       0.17   & 0.67 & 3.2 & 40 & 1.03  & 1 & 1.6 & 20 \\
       0.08 & 0.3 & 3.2 & 40 & 0.78 & 0.75  & 1.6 & 20 \\
       0.03 & 0.17  & 4 & 50 & 0.52 & 0.5  & 1.6 & 20 \\
       0.01 & 0.083  & 4.8 & 60 & 0.37   & 0.8  & 2.4 & 30 \\
       0      &    0     &  2.4 & 30 & 0.23   & 0.5  & 2.4 & 30 \\
       ~&~&~&~&0.21 & 0.2  & 1.6 & 20 \\
       ~&~&~&~&0.15 & 0.6  & 3.2 & 40 \\
       ~&~&~&~&0.08 & 0.3  & 3.2 & 40 \\
       ~&~&~&~&0.03 & 0.2  & 4 & 50 \\
       ~&~&~&~&0 & 0 & 2.4 & 30 \\  \hline
  \end{tabular}
  \caption{Experimental parameters, Set 1: 3-5 mm bubbles generated using capillary needles of inner diameter of 500 $\mu$m, Set 2: 2-4 mm bubbles produced using capillary needles of inner diameter of 120 $\mu$m} 
  \label{tab:kd}
  \end{center}
\end{table}

We obtain $b$ parameter values from $\infty$ to $0$ by varying the void fraction ($2$ to $0\%$) and the mean flow velocity ($0$ to $60$ cm s$^{-1}$). Table ~\ref{tab:kd} lists all the different parameters varied in the present experiments.  The turbulent flow properties (e.g. $u^{\prime}_0$) are characterised by combined CTA - LDA (Laser Doppler Anemometry) measurements of only the liquid phase at different mean flow speeds (for details see~\cite{prakash2}). The Kolmogorov scales range from 410 $\mu m$ to 200 $\mu m$ for mean flow speeds ranging between 20 cm/s to 60 cm/s respectively. For the same range of mean flow speeds, the Taylor length scales range between 10 $mm$ to 6 $mm$, and the integral length scales are about $\sim$60 $mm$.

\subsection{High-speed imaging}

\begin{figure}
\centering
\includegraphics[width=14cm]{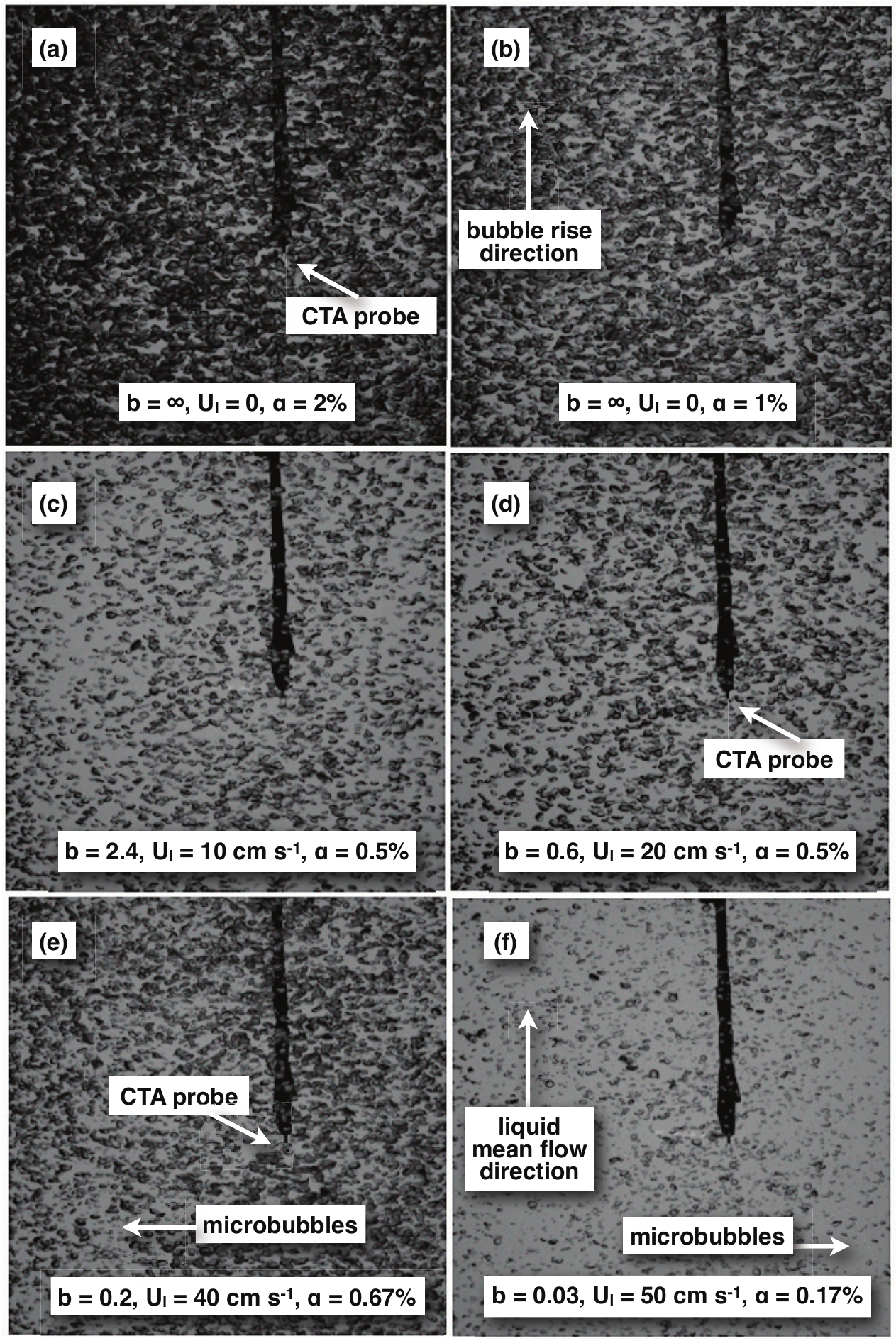}
\caption{Snapshots from the experiments (set 1, see Table~\ref{tab:kd}) at different conditions.}
\label{fig:exp_snapshot}
\end{figure}

In order to visualise the flow, a Photron-PCI 1024 high-speed camera was focused on a vertical plane at the centre of the measurement section. We acquired two-dimensional images of each experiment using the camera (at $1000 Hz$) and some of these snapshots are shown in Figure~\ref{fig:exp_snapshot}. The $b=\infty$ experiments are shown in Figure~\ref{fig:exp_snapshot}(a) and (b) where the gas void fractions are $\alpha$ = 2$\%$ and 1$\%$, respectively. The dense nature of the flow at such void fractions is evident: the flow is opaque and the phase-sensitive CTA probe is barely visible. As we proceed to look at the other cases in Figure~\ref{fig:exp_snapshot}(c)-(f), $\alpha$ decreases, and the liquid mean flow speeds ($U_{l}$) increase, corresponding to a decrease in the $b$ parameter from $\infty$ to $0.03$. In our experiments, the bubbles must pass through the active-grid, which consists of randomly oscillating steel flaps ($\approx$ few rotations per second) to generate the required turbulence. At any given instant in time, the active grid is $50\%$ transparent (open) to the flow. Hence, the bubbles face a slight obstruction and sometimes interact with the steel flaps. The bubble-flap interaction, however, causes fragmentation of the bubbles and results in a decrease of the diameter of the bubbles. This bubble diameter decrease becomes apparent at higher liquid mean flow speeds ($U_{l}$), as seen in Figure~\ref{fig:exp_snapshot}(c)-(f). In our experiments, the CTA probe is placed sufficiently far away ($\approx1m$) from the turbulence generation source (active grid) - so that the large-scale eddy motions and other transients  have decayed and the turbulence is almost homogenous and isotropic at the location of the probe. The probe location is at a distance ($1m$) where the downstream root-mean-square stream-wise velocity fluctuation does not vary significantly (see~\cite{Poorte2002} for a comparison between the present active-grid and the more traditional passive grids). 

\begin{figure}
\centering
\includegraphics[width=11cm]{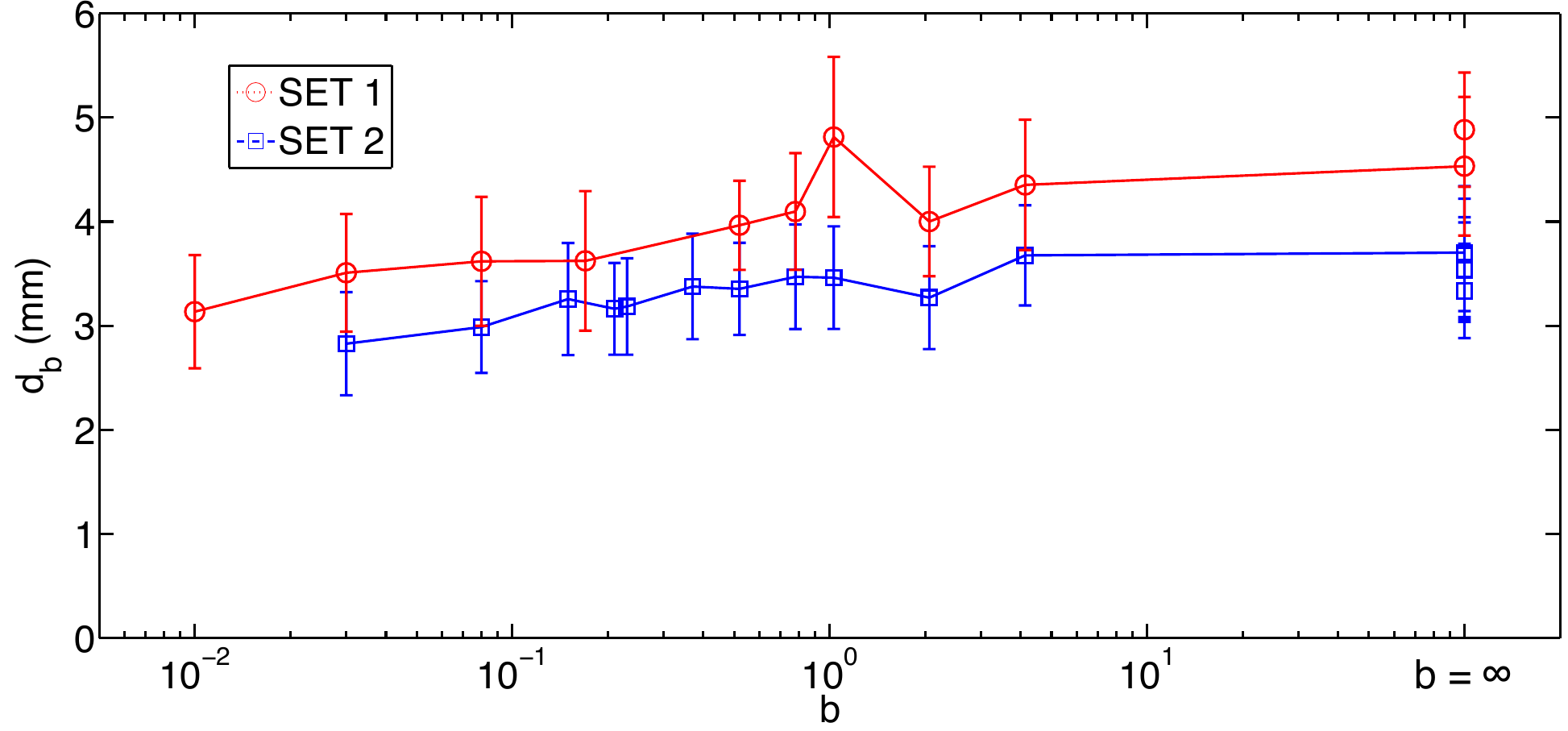}
\caption{Bubble diameter versus the $b$ parameter. The error bars are estimated based on the standard deviations.}
\label{fig:sizevsb}
\end{figure}

We obtain a quantitative measurement of the bubble diameter using the images acquired (Figure~\ref{fig:exp_snapshot}) from the individual experiments. The bubbles highly deform over time, and given the dense nature (high $\alpha$) of the flows, there is currently no reliable automated image processing algorithm available to accurately determine the bubble diameter. Hence, we had to resort to a manual procedure - where individual bubble boundaries are marked using mouse-clicks in the open-source ImageJ software. An ellipse is fitted to the deformed bubble boundaries, and the equivalent bubble diameter is calculated as: $d_b = \sqrt[3]{d_{l}^2d_{s}}$, where $d_l$ and $d_s$ are the long and short axes of the ellipsoidal bubble. In each experiment, we measure the diameters of $\approx50-100$ bubble samples, and then take the mean value of the distribution to be the equivalent diameter of the bubble. In Figure~\ref{fig:sizevsb} we observe that the bubble diameter decreases with a decrease in the $b$ parameter (increasing liquid mean flow speeds). The decrease in the bubble diameters at higher liquid mean flow speeds is mainly due to the fragmentation of the bubbles (as described above) (also see~\cite{prakash1}). 
 Here, the error bars represent the standard deviation of the measured distribution of bubble diameters.  

At high mean flow speeds, air is entrained from an open vent at the top of the TWT because of oscillations of the free-surface exposed to the atmosphere. The entrained air unavoidably results in micro-bubbles which are fed back into the measurement section and contaminate the flow. These entrained micro-bubbles pose a problem at mean flow speeds higher than $30$ cm s$^{-1}$, and are visible (as very small bubbles) in Figure~\ref{fig:exp_snapshot}(e) and (f). It is necessary to account for these micro-bubbles in the data analysis, and this issue will be discussed further below. 

Also, Figure~\ref{fig:exp_snapshot}(e) and (f) show instantaneous snapshots of the bubbles at a particular instant of time. Although there might be localised clustering or non-homogeneity of the swarm of bubbles in different regions, these are dynamic and change continuously over time. The homogeneity of the gas phase decreases at low gas volume fractions - this is an inherent limitation in the experiments.

The present experiments in the pseudo-turbulence regime ($b=\infty$) for set 1, are essentially the same as the measurements carried out in~\cite{julian2}, and serve as a reference case for the data analysis and results. For this case of freely rising bubbles in a quiescent liquid, the bubble-based Reynolds number is $Re = d_b U_r / \nu \approx 1000$~\cite[]{julian2}, where
$\nu$ is the kinematic viscosity of water ($1\times10^{-6}$ m$^2$ s$^{-1}$). %Weber number $We \approx 3$, and Eotvos numbers $Eo \approx 3$     

\subsection{\label{sec:level2}Phase-sensitive constant temperature anemometry}

Hot-film anemometry is an important technique in single-phase turbulent flows, but its application in bubbly flows is not straightforward. Since it is an intrusive technique, bubble-probe interactions result in disturbances in the hot-film time series voltage signal. Various methods have been developed in the past to remove these `bubbly spikes'~(\cite{zenit,judith,julian}), so as to exclusively analyse only the liquid fluctuation segments measured by the hot-film probe. For example, a threshold method was used by~\cite{zenit} and \cite{julian} and a pattern-recognition method was used by~\cite{judith}. These methods essentially come up with an indicator function that labels the gas and liquid phase separately. However, a much better approach to eliminate the bubbly spikes from the CTA signal is to measure the indicator function \textit{in-situ} during the experiments. This can be done by attaching optic fibres with a diameter of $\sim$ 100 $\mu$m close to the hot-film probe (at a distance $\sim$1 mm) to detect the gas phase. Light is continuously passed through the optic fibre, and when a bubble  collides with the probe, the change in refractive index of the gas phase results in a signal change. This technique, called the phase-sensitive Constant Temperature Anemometry (CTA) was developed by~\cite{ramon2011} and is used in~\cite{julian2} and~\cite{mendez2013}. This method can be used to directly detect and remove the bubbly spikes in the hot-film signal.

In this work, we follow almost the same experimental procedure and analysis as in~\cite{julian2}, but the important difference here is that we vary the $b$ parameter over a wide range to cover the regimes between pseudo-turbulence ($b=\infty$) and single-phase turbulent flow ($b=0$).  The phase indicator function obtained using information from the optic fibre signal labels the liquid fluctuations and bubble collisions separately. This is used to remove the bubbly spikes and separate the segments containing only liquid fluctuations from the time series signal for further analysis. The power spectrum was calculated for each segment of liquid fluctuation and averaged to obtain the spectrum for a particular case of $b$. The phase-sensitive CTA probe (DANTEC cylindrical probe - 55R11) is calibrated by simultaneous measurement of absolute velocities of the single-phase using a DANTEC Laser Doppler Anemometry (LDA) setup (as in~\cite{prakash1,prakash2}). The standard King's law fit is used for the voltage-velocity data. The acquisition rate was 10 kHz and the measurements were carried out for durations of 30 min, and we repeated the experiment once again. We average the results from the two experiments for each case of $b$. The CTA-LDA calibration was carried out separately, and once the calibration was done, we did not disturb the probe or experimental setup and we then took measurements of the different cases of $b$. We kept the duration of experiments short (30 min) to avoid problems with the calibration.

The CTA probe is a fiber-film probe that is suitable for use in liquids, it is more robust than wire probes and less sensitive to contamination. The fiber sensor is a 70 $\mu m$ diameter quartz fiber, with an overall length of 3mm,  and covered by a nickel thin film approx. 0.1 $\mu m$ in thickness. The rated minimum and maximum velocity are 1 cm/s and 1000 cm/s, and the rated maximum frequency limit is 30 kHz. The size of the sensitive area of the hot-film is 1.25 $mm$, and its response time is 0.033 $ms$. The smallest Kolmogorov time-scale of the single-phase turbulence is 35 $ms$ (for a mean flow speed of 65 cm/s), hence, the probe response time is well within the Kolmogorov time scale. Also, the Kolmogorov length scales range from 410 $\mu m$ to 200 $\mu m$ for mean flow speeds ranging between 20 cm/s to 60 cm/s respectively. The sensitive area of the CTA probe (1.25 $mm$) will be able to resolve the Kolmogorov scales. The CTA probe is placed such that the sensitive area is longitudinal or against the flow direction. The optical fibres are attached in such a way that they are anti-parallel to the direction of gravity and their tips oppose the rising bubbles. The optic fibres are carefully attached to the CTA probe using permanent glue before placing it in the water tunnel. 
It has been previously established that the presence of the optic fibres does not compromise the probe's bandwidth and that its infuence on the power spectrum is negligible (\cite{ramon2011}).

The phase-sensitive CTA technique works very well for pseudo-turbulent bubbly flows where the bubble diameters are in the range $\sim$ 2-5 mm. However, when micro-bubbles collide with the CTA probe, the optic fibres will not be able to register the collision. The reasons for this are two-fold: (i) the micro-bubbles are small in size ($\lesssim300$ $\mu$m diameter), and (ii) the separation distance between the CTA probe and the optical fibres is larger ($\sim1$ mm) than the micro-bubble size. As we mentioned before, micro-bubbles cause a contamination in the present experiments when the mean flow speeds exceed $30$ cm s$^{-1}$. In these experiments, we inevitably use a threshold method to remove the micro-bubble collisions, in addition to the phase information obtained from the optic fibres. Further, to keep the data analysis consistent, the combination of the optical fibres and the threshold method is also used in all the experiments except pseudo-turbulence, where only the optical fibres are used.

\subsection{Spectrum calculation}

\begin{figure}
\centering
\includegraphics[width=13cm]{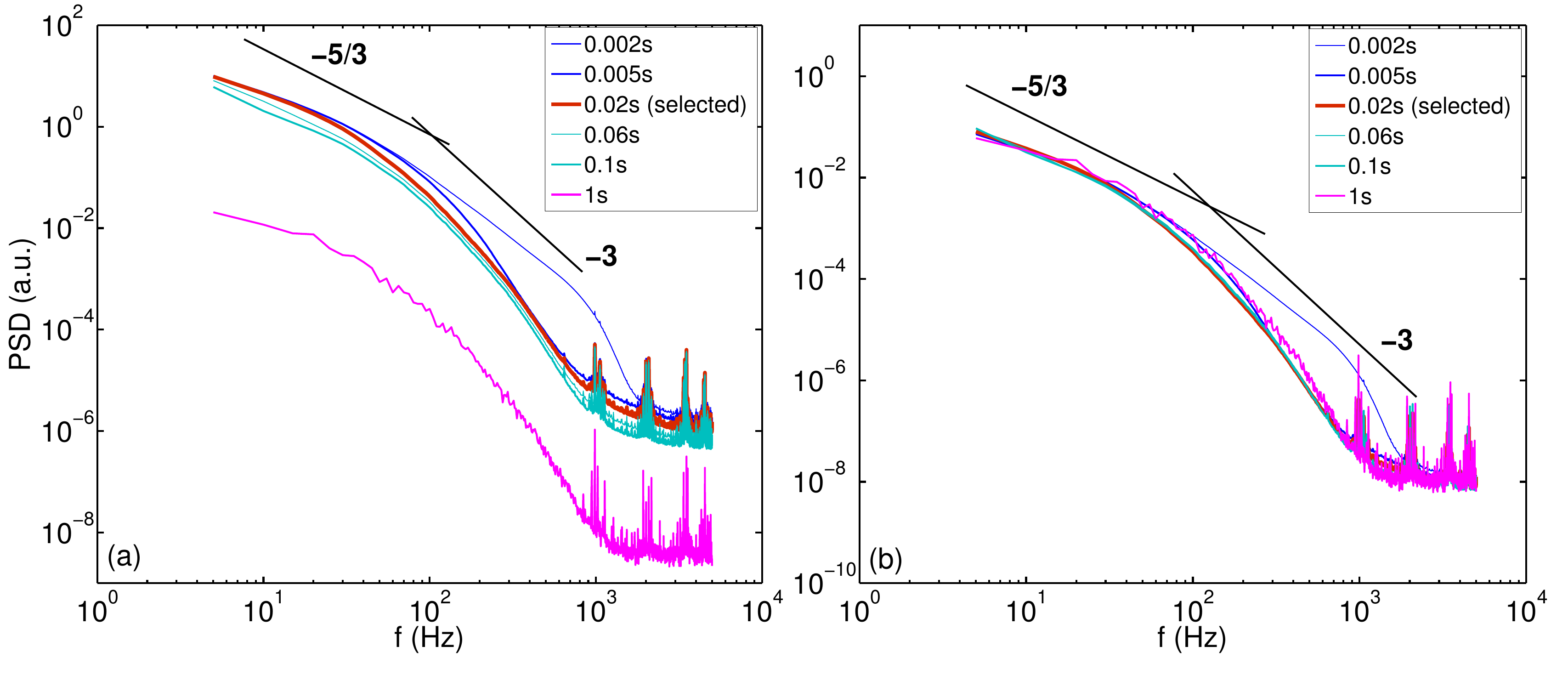} 
\includegraphics[width=9cm]{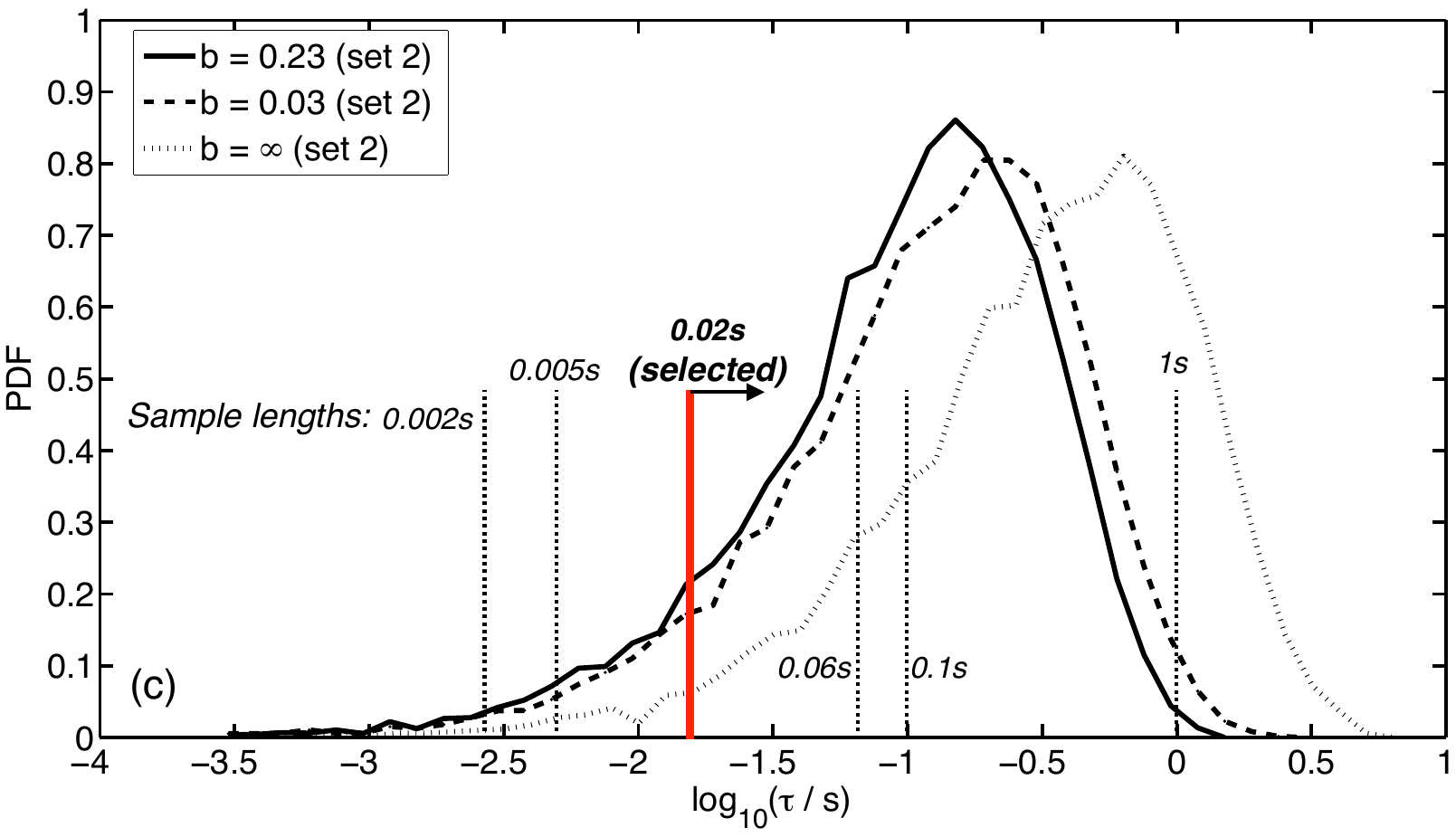} 
\caption{Minimum sample duration criterion for the calculation of the energy spectrum, for the case $b$ = 0.23 (set 2) (a) non-normalised spectra (b) normalised spectra. The different coloured lines are the spectra obtained with various sample durations, as shown in the legend. The normalised spectrum (b) does not change when the sample duration is sufficiently long. A sample duration of $0.02$s is found to be optimal in the present work where the sampling frequency is 10 kHz. (c) The PDF of the sample duration used to calculate the spectrum for the cases $b$ = $0.23, 0.03, \infty$ (set 2). The solid
red vertical line corresponds to the selected sample duration of $0.02$s; and it is evident (right side of the red line) that more than $\sim90\%$ of the signal is considered for spectrum calculation  in all the three cases.}
\label{fig:samplesize}
\end{figure}

We now consider the  spectral distribution of the liquid velocity fluctuations in the different regimes of b. As mentioned in the introduction,  the energy spectrum, or more precisely, the  Power Spectral Density  (PSD) of the velocity fluctuations were  reported  by many authors both for  pseudo-turbulent  and for bubbly turbulent flows. The PSD forms therefore a convenient  means to  describe these flows. We start by explaining  how the PSDs were obtained in our case. The PSD was calculated for individual segments of the liquid fluctuations (free from bubbly spikes) using the Welch method (using hamming windows) at fixed frequencies, and then averaged over all the liquid segments in the measurement to obtain the final result (as in~\cite{julian2}). The segments selected for the spectrum calculation must have a certain minimum duration to properly resolve all the frequencies; segments that are too short will lack information on the large length-scales (low frequencies) and will simply add to the high frequency components of the spectrum (noise). We investigate the effect of varying this minimum sample duration (in seconds) criterion on the spectrum for a selected case of $b=0.23$ (set 2); the results are shown in Figure~\ref{fig:samplesize}. We observe drastic changes in the spectra depending on the value of the minimum sample duration. In Figure~\ref{fig:samplesize}(a) we show the spectra directly obtained from the calculation. Both the amplitude and the scaling change; with a monotonic decrease in energy with increase in sample duration. This is expected because when we increase the minimum sample duration considered, we have fewer segments considered and the average energy decreases. In the extreme case of minimum sample duration of $1$s, the spectrum looks noisy as it was averaged only over 80 segments. For the results presented here, we have selected an optimal minimum sample duration of $0.02$s, which nicely resolves all the frequencies (the spectrum is averaged over 14500 segments). In Figure~\ref{fig:samplesize}(b), we show the same spectra after normalising the area under the curve to be equal to unity. We observe that the selected value of $0.02$s is an optimal value; the extreme values ($0.002$s, $1$s) show deviations in the scaling. In addition, Figure~\ref{fig:samplesize}(c) shows the PDF of the 
of the logarithm of the sample durations for three different cases $b$ = $0.23, 0.03, \infty$ (set 2). We observe in all cases that more than $\sim90\%$ of the samples have a duration larger than our chosen minimum sample duration of $0.02$s; hence our PSD calculations are robust.  
In the sections that follow, we present results for the normalised spectra in all the cases of $b$ parameter as it allows us to focus solely on changes in the scaling. 
The energy in the non-normalised spectra depends on the number of samples considered in the averaging, and this can differ in each experiment as it depends on the flow conditions; hence, we normalise the spectra.

%%%%%%%%%%%%%% END EXPERIMENTS SECTION

%%%%%%%%%%%%%%% RESULTS SECTION BEGINS

\section{Results}

\subsection{\label{sec:level2}Liquid velocity statistics}

\begin{figure}
\centering
\includegraphics[width=12cm]{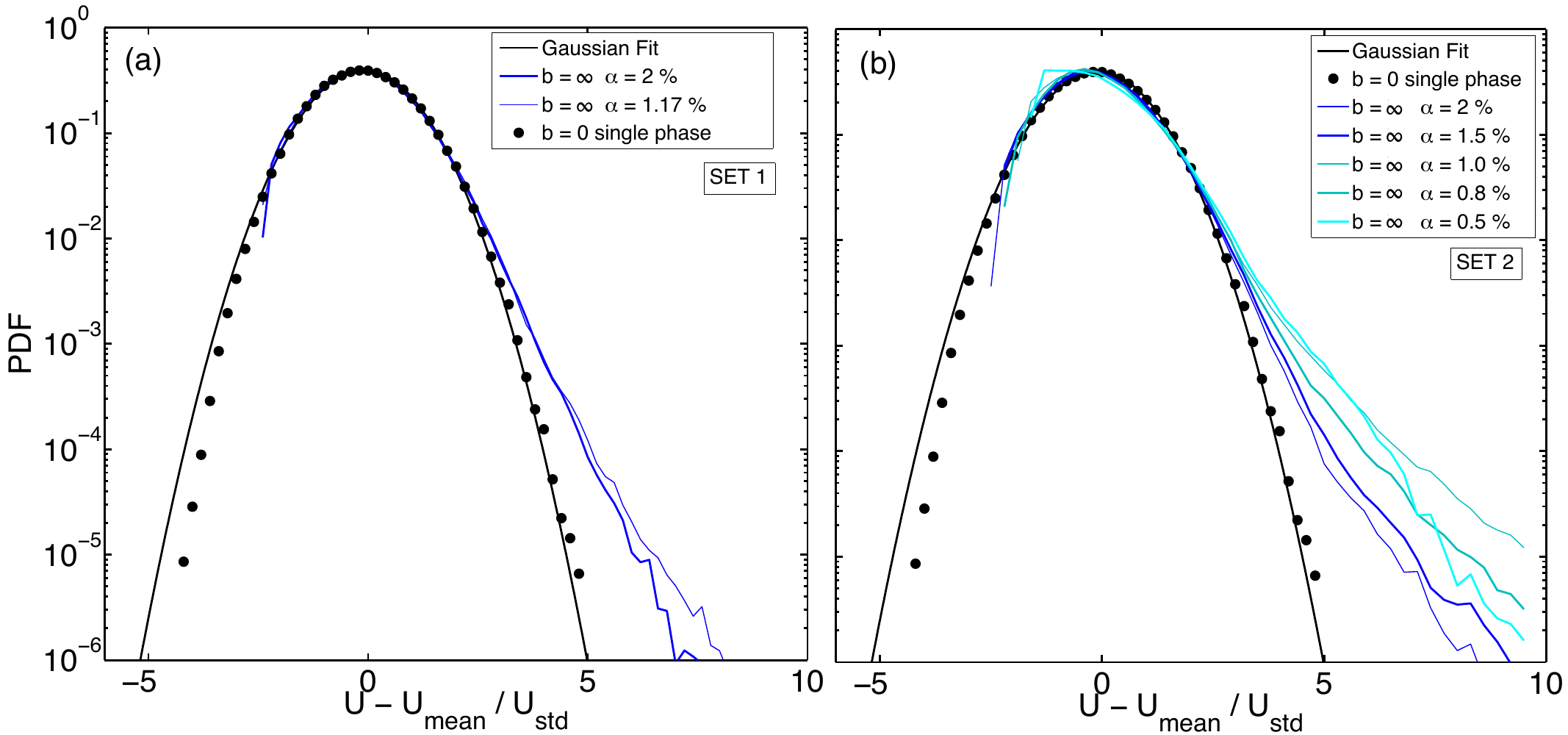}
\includegraphics[width=12cm]{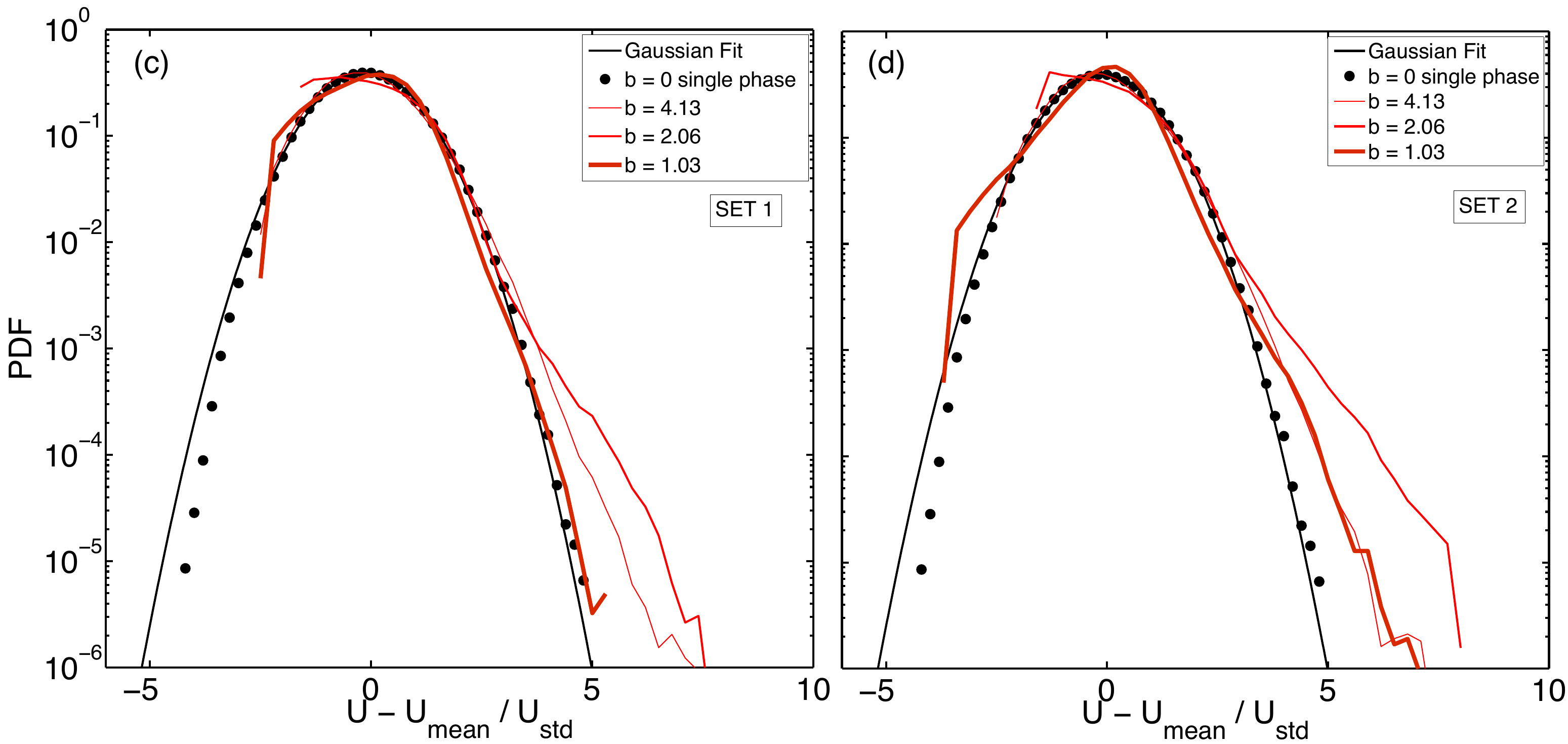}
\includegraphics[width=12cm]{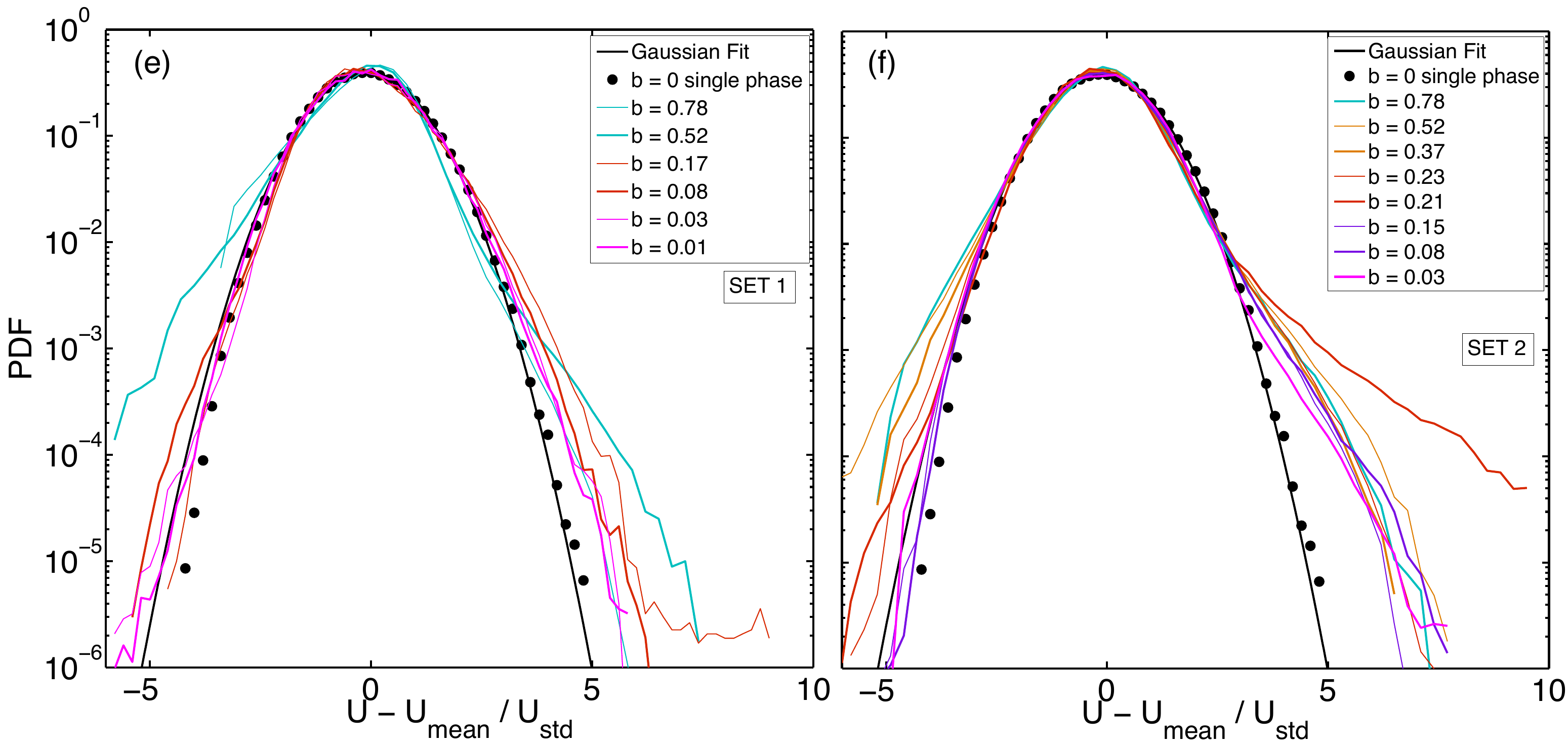}
\caption{The liquid velocity normalized by the standard deviation $U_{std}$, PDFs for different $b$, left panel: set 1, right panel: set 2. (a, b) represent $b = 0, \infty$, (c, d) $b = 0$ and $> 1$, (e,f) $b = 0$ and $< 1$. All the bubbly flow cases show deviations from the Gaussian profile with enhanced probability of upward fluctuations.  }
\label{fig:velpdf}
\end{figure}

The vertical component of the liquid velocity time-series was measured using the phase-sensitive CTA technique at different $b$ parameter values. We now consider the statistics of these liquid velocity fluctuations using the separated segments of the signal which are free from the bubble collisions.    
In Figure~\ref{fig:velpdf}, we present the normalized liquid velocity probability density functions (PDFs) for the different values of $b$ covered in the present  work, including both set 1 and set 2 experiments (see Table~\ref{tab:kd}). The liquid velocity PDF for single-phase turbulent liquid ($b=0$) at a mean flow of 30 cm s$^{-1}$ (Taylor Reynolds numbers $Re_\lambda = 170$) (black dots) nearly follows Gaussian statistics. This single-phase result serves as the reference case.

The liquid velocity PDFs for the cases with bubbles (b $>$ 0) are asymmetric and show a deviation from Gaussian behavior. The positive tails of the PDFs show higher probability compared to the Gaussian profile, originating from the flow entrainment in the wake of the rising bubbles, which leads to a larger probability of upward fluctuations~(\cite{risso,riboux}). 

In Figure~\ref{fig:velpdf} there is no clear trend w.r.t. the $b$ parameter in the PDFs, the only consistent observation is the asymmetry in the PDFs when bubbles are present (b $>$ 0). Also, in Figure~\ref{fig:velpdf}, for each row comparing set 1 and set 2 experiments, the PDFs for two similar values of $b$ parameter do not necessarily collapse. Hence, the $b$ parameter does not seem to be a satisfactory single parameter to characterise velocity statistics in different regimes of turbulent bubbly flow. The positive asymmetry of the PDFs is highest for pseudo-turbulence ($b = \infty$) and the asymmetry gradually reduces at smaller $b$ parameter values probably due to the increase in turbulence levels.

\subsection{\label{sec:level2}Energy Spectra}

\begin{figure}
\centering
\includegraphics[width=14cm]{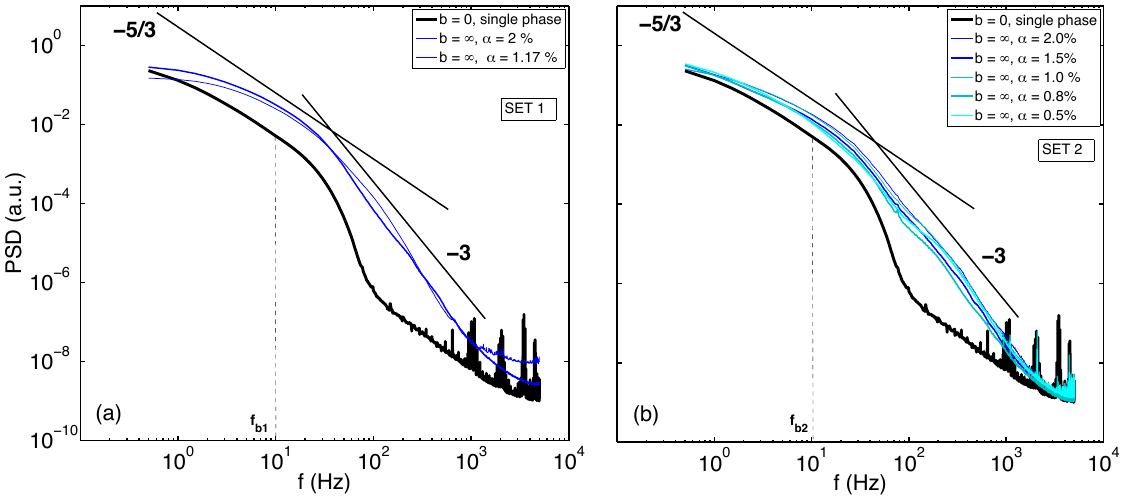} 
\includegraphics[width=14cm]{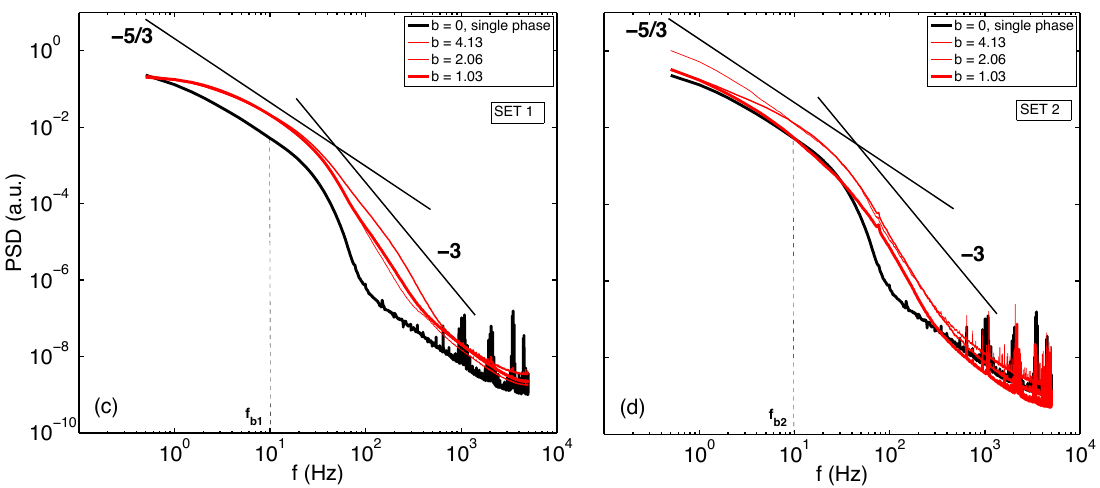} 
\includegraphics[width=14cm]{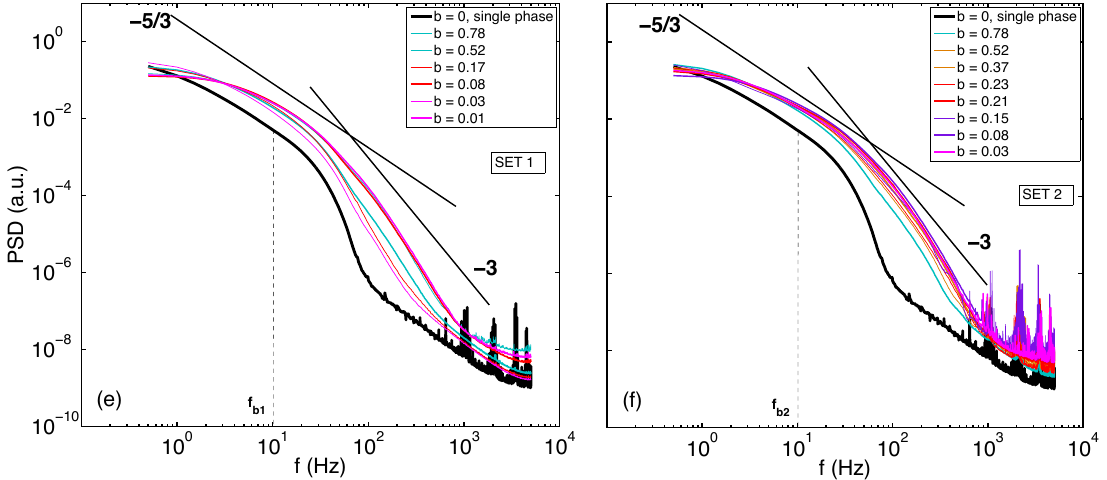} 
\caption{The normalised energy spectra at different $b$, left panel: set 1, right panel: set 2. (a, b) represent $b = 0, \infty$, (c, d) $b = 0$ and $> 1$, (e,f) $b = 0$ and $< 1$. All the bubbly flow cases show deviation from the {\it{-5/3}} Kolmogorov single-phase spectrum beyond the transition frequency (at 10 Hz as indicated with the dotted line), and reasonably follow the {\it{-3}} scaling.
}
\label{fig:spectrum}
\end{figure}

\begin{figure}
\centering
\includegraphics[width=12cm]{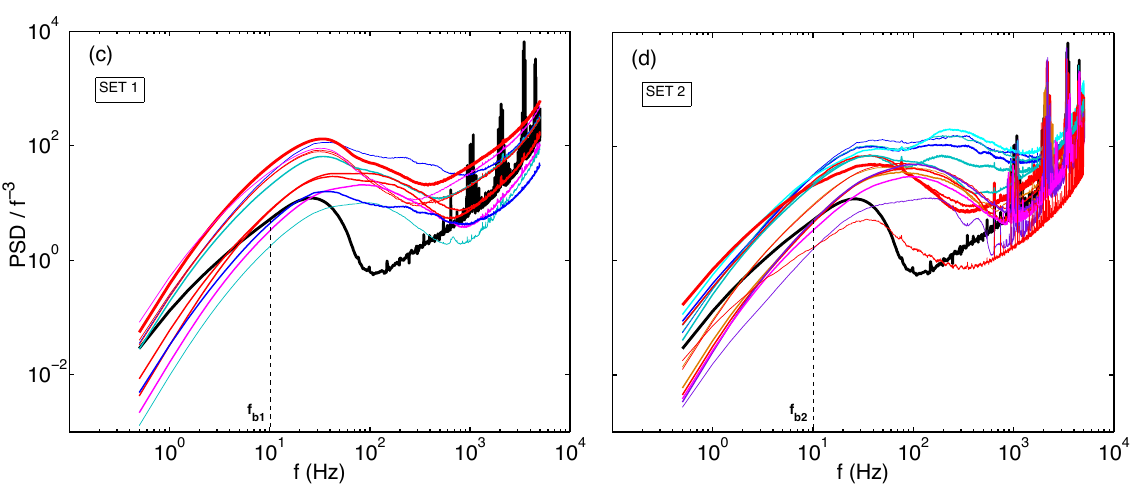}
\includegraphics[width=12cm]{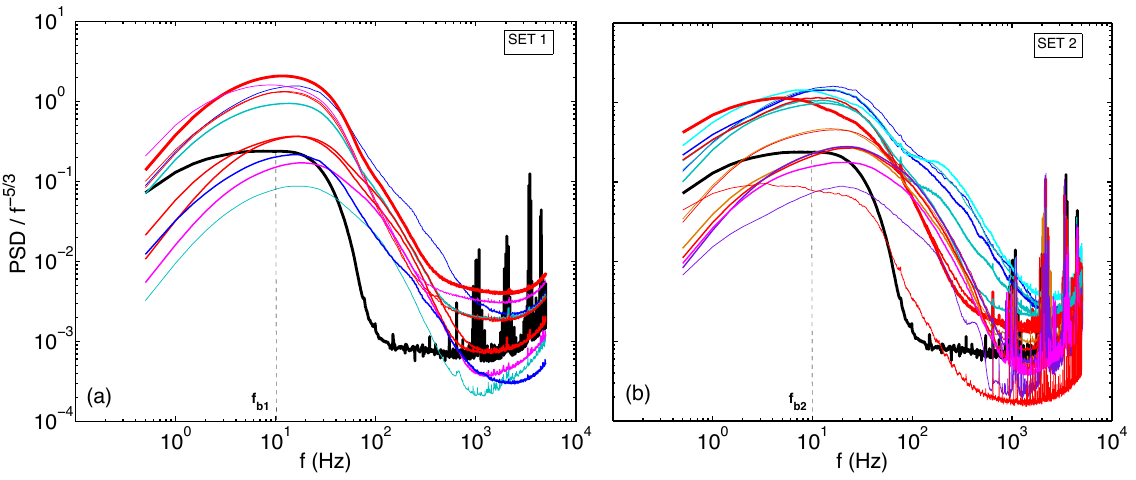}
\caption{The compensated energy spectra at different $b$. left panel: set 1, right panel: set 2. (a,b) spectra compensated with {\it{-5/3}}, (c,d) spectra compensated with {\it{-3}}. The thick black lines indicate the single phase case. The colours for the bubbly flow cases correspond to the scheme in Fig~\ref{fig:spectrum}. As observed in (c,d), all the bubbly flow cases roughly exhibit the {\it{-3}} scaling.}
\label{fig:spectrum_compensated}
\end{figure}

In figures~\ref{fig:spectrum} and \ref{fig:spectrum_compensated}, the PSDs for different  values of  b  are shown.  Figures~\ref{fig:spectrum}a-~\ref{fig:spectrum}c, ~\ref{fig:spectrum}e show PSD for the experiments  in set 1, as given in detail in table~\ref{tab:kd}, and likewise  figures~\ref{fig:spectrum}b,~\ref{fig:spectrum}d, \ref{fig:spectrum}f for set 2.   Frequency is plotted   along the horizontal axis in each of the figure~\ref{fig:spectrum}.  On the vertical axis is the PSD per frequency in arbitrary units but such that the area under the  curve in question equals unity.  As landmarks are shown in all of figure~\ref{fig:spectrum}: the single phase Kolmogorov -5/3  spectrum, and the lines with slope $-$5/3  and $-$3 respectively. In figures~\ref{fig:spectrum}a and \ref{fig:spectrum}b all results for the pseudo-turbulent  case b $= \infty$, but  different values of the concentration  by volume fraction $\alpha$, are collected. They all clearly follow the $k^{-3}$ line. Figures~\ref{fig:spectrum}c,d show the results for b $>$ 1, whereas figures~\ref{fig:spectrum}e,f show the results for  0 $<$ b $<$ 1.  From the definition of the bubblance parameter b, measuring the relative importance of the bubbly energy  content with respect to the turbulent energy, one may  expect  dominance of the bubbles for the larger b, let's say b $>$ 1, and therefore  k$^{-3}$ type of spectra, and dominance of $k^{-5/3}$ for small values of  b (b $<$ 1). The first part of this expectation is true, as figures~\ref{fig:spectrum}c,d  show. The curves for various  b  values do not exhibit much differences, except  that   in the higher frequency range, of about  100 Hz,  there is a slight  decrease of energy with increasing  b.  The second part of the expectation ($k^{-5/3}$ for small values of  b), however, is not realized. Surprisingly, even for very small  b, the PSD curves have  a slope  close to $-$3. Apparently  even a small concentration by bubbles alters the PSD significantly and changes  it into a  PSD much like  in the pseudo-turbulent case. The same  trend shows in the compensated spectra in figure~\ref{fig:spectrum_compensated}. In figures~\ref{fig:spectrum_compensated}c,d, it appears that the $k^{-3}$ scaling is followed fairly well.

\section{Interpretation of results, discussion and comparison with other work}

A better understanding of  the results is obtained when we consider a frequency that is representative for the bubbles.  Such a frequency is
\begin{equation}
 f_{b} = \frac{U_r}{2\pi d_{b}}.
 \label{eq:freq}
\end{equation}
 The representative bubble frequencies here are: $f_{b1} \sim9$ $Hz$ and $f_{b2} \sim12$ $Hz$ when the bubble diameters are in the range $d_{b1} \sim$ 5-3 $mm$ (set 1)  and $d_{b2} \sim$ 4-2 $mm$ (set 2) respectively. Since $f_{b1}$ is close to $f_{b2}$, we consider $f_b \sim 10$ Hz.
Now we draw in figures~\ref{fig:spectrum}a-f  a vertical line at the  value of $f_b$ ($\sim10$ Hz) pertaining to the two sets of experiments. 
 Doing this we see, as shown in figure~\ref{fig:spectrum}, an  interesting  phenomenon:  This line  marks the transition  of a part of the PSD, to the left  of the line, where the $-$5/3 slope is followed, and a part to the right, where the $-$3 slope is followed.  Energy is fed in the fluid by the bubbles at frequencies close to $f_{b}$.  This is accompanied by the production of eddies of the size of the bubbles.  According to \cite{lance} ``The eddies are thus dissipated by viscosity, before spectral transfer can take place".  This explains  the separation of the PSD in two parts as described above.  Since the eddies do not take part in the large scale energy transfer, the spectra follow the  Kolmogorov  $-$5/3  slope in the frequency range  between zero and  $f_{b}$.  The energy input by the bubbles passes over only to higher wave numbers.  In the steady case, there is an equilibrium between the energy production and the dissipation.  In Fourier space the  dissipation  is $\nu E(k)k^2$, where $E(k)$ is the Fourier transform of the kinetic energy, divided by density, and $\nu$ the kinematic viscosity of the fluid. The power input  by the bubbles  in a uniform bubble distribution is  independent of the space coordinate and equal to the work done by buoyancy, which is $\alpha gU_r$ divided by density. Since this is constant, its Fourier transform is $\alpha gU_r/k$, and hence the equilibrium between energy input and energy dissipation in (one-dimensional) Fourier space is
\begin{equation}
\nu E(k)k^2 \sim \frac{\alpha g U_r}{k}.
\label{eq:ek}
\end{equation}
This leads to  a behavior $E(k) \sim k^{-3}$, as shown in our experiments.  Of course this  regime does not go on into the dissipative  Kolmogorov scale ($\eta_K$).  We obtained eq.~(\ref{eq:ek})  in much the same way as \cite{lance} did, with the difference that we  used the Fourier transform of the power input. 
%they obtained eq.~\ref{eq:ek} on dimensional grounds, whereas

At the start of our experiments reported in the present paper, we assumed that the ÒbubblanceÓ parameter defined  in (1.1), would be convenient in marking the transition from $k^{-5/3}$ to $k^{-3}$ behavior of the spectra. The results indicate that it is rather the bubble frequency, defined in (4.1), which is useful as indicator of the transition. This however, opens up the question of the mechanism producing this transition. In particular, the question of how only a few bubbles are sufficient to bring about the $k^{-3}$ spectral scaling behavior needs to be investigated further. It is clear that the transition is due to phenomena on the scale of the bubbles. These include the individual bubbles and their wakes, as well as interaction between them. The fluctuations in the experiments by Lance $\&$ Bataille  (1991) were solely caused by the rising bubbles. Lance $\&$ Bataille  (1991) made the hypothesis that the $k^{-3}$ behavior of the spectra is due  to the fact that eddies produced by bubbles are dissipated before they can take part in the energy cascade.
%Our experiments show that this  is also the case when the bubbles rise in turbulent flow.}

The $k^{-3}$ spectrum scaling is a very robust characteristic of pseudo-turbulence as shown in many experimental results, e.g. \cite{riboux,julian2,mendez2013}. Here it is shown  that also in bubbly turbulent flow this is the case at frequencies above $f_b$ defined in eq.~\ref{eq:freq}, independent of the bubble concentration.  Also theoretically the  $k^{-3}$   slope in pseudo-turbulence has been proven to  be robust, because derivations were obtained along different lines.  \cite{risso2011} and \cite{Amoura} distinguished between  spatial and  time  fluctuations  of the velocities at a particular point in the bubbly flow, and he found the $k^{-3}$ spectral density for both spatial and  temporal parts. Of particular relevance for the present work is the numerical simulation by \cite{riboux2013}, in which they considered the flow through a random array of fixed bubbles.  Let us say, the mean flow is  downward. If we impose an upward velocity on both the fluid and the bubbles, the situation is that of bubbles rising with constant  velocity upwards. The bubbles are modelled by fixed momentum sources randomly distributed, and this flow is very near to what is assumed here in deriving eq.~(\ref{eq:ek}). \cite{riboux2013} numerically solved the Navier-Stokes equations, and the results for PSDs are shown in their figures 7 and 8 as function of the reciprocal wavelength $\lambda^{-1}$.  For the simulations reported in their Figures the bubble diameter is  2.5 mm, for which the terminal rise velocity $U_r$ is about 0.3 m/s. They exhibit indeed in a range of  $\lambda^{-1}$ values, between small and large ones, the $k^{-3}$ slope. It is interesting that  at the low $\lambda^{-1}$  side the slope starts to show $k^{-3}$  behavior at a value of  $\lambda^{-1}$  of about $10^2$ m$^{-1}$. Using  $U_r= 2\pi f/k$, one can convert $\lambda_s^{-1}=k_s/2\pi$   into a starting frequency of $f_s \sim 30$ Hz for the {-3} scaling. The frequency  $f_b$ representative for bubbles of diameter 2.5 mm would according to (4.1) be  $0.3/(2\pi*2.5* 10^{-3}) Hz = 19$ Hz, which is of the same order of magnitude. This indicates that the {-3} spectrum scaling first shows up in their simulations also at frequencies higher than the frequency given in (4.1) associated with the bubble size in their simulations. Of course this correspondence  is only qualitative and one should be cautious to attach further extrapolations to it, bearing in mind as well that in their numerical model there is no external turbulence as in the present study. 

%In the case of these figures the bubble diameter is 2.5 mm, and for these bubbles the terminal rise velocity $U_r$ is  about  0.3 m/s.  They exhibit  indeed in  a range of wavenumbers, in between large and small ones, the $k^{-3}$ slope.  But  what   is particularly interesting, is that at the low wave number side  the slope starts to take $k^{-3}$   behavior  at a wavenumber $k_s$ of about  $10^2$. Using $U_r=2\pi f/k$, one can convert $k_s$  into the starting frequency for the $-$3 scaling at $f_s=5 s^{-1}$ for their spectra. 
% The frequency $f_b$ that is representative for the bubble scale according to eq.~(\ref{eq:freq}) would be  $0.3/(2\pi \times 2.5 \times 10^{-3}) = 19$ $s^{-1}$, which is in the same order with $f_s$. This indicates that the $-$3
% spectrum scaling also shows up at the scale smaller than the corresponding size of the bubbles in their simulations. One should however be cautious to attach further extrapolations to this correspondence because in the  numerical model there is no external turbulence as in the present study.

\section{Summary}
In this work, we studied the energy spectra of the velocity fluctuations in different regimes of turbulent flows with varying bubblance parameters (b): single-phase turbulence (b = 0), turbulence with some bubbles (0 $<$ b $<$ $\infty$) and pseudo-turbulence (b = $\infty$). With varying b, one may expect to find a gradual change in the spectral slope from $-3$ ($b = \infty$, pseudo-turbulence) to $-5/3$ ($b = 0$, single-phase turbulence). However, the results do not show a smooth transition: the $-$3 spectrum scaling was found at length scales smaller than the size of the bubbles even for very small $b$ parameter values (b $\sim 0.01$), i.e. when the void fraction is as low as 0.1$\%$. Hence, the bubbles are able to modify the spectra very efficiently, even though they are surrounded by the external turbulent fluctuations. Following the argument by \cite{lance}, we provided an explanation on the $-$3 spectrum scaling by balancing the energy production of the bubbles with the viscous dissipation in the Fourier space. The $-$3 spectrum scaling seems to be a generic feature of turbulent bubbly flows. Regarding the bubblance parameter, in future work, it would be a worthwhile effort to come up with a more suitable parameter to characterise turbulent bubbly flows.

\section*{Acknowledgments}

\smallskip
We thank Andrea Prosperetti, Roberto Zenit, Dennis van Gils, Dennis Bakhuis, and Varghese Mathai for beneficial discussions. 
%We also thank F. Ernesto Mancilla Ramos for assistance with the experiments. 
We are grateful to Gert-Wim Bruggert, Martin Bos and Bas Benschop for their continued support. We acknowledge support from the Foundation for Fundamental Research on Matter (FOM) through the FOM-IPP Industrial Partnership Program: \emph{Fundamentals of heterogeneous bubbly flows}. We also acknowledge support from  the European High-performance Infrastructures in Turbulence (EuHIT) consortium. CS acknowledges the support of NWO by VIDI Grant No. 13477. Finally, we thank the anonymous referees for their constructive comments and suggestions which improved the manuscript.

\bibliographystyle{jfm.bst}

\end{document}